\documentclass[letterpaper]{jpconf}
\usepackage{graphicx}
\begin{document}
\title{The PIENU experiment: a precision measurement of the $\pi\rightarrow e\nu / \pi\rightarrow\mu\nu$ branching ratio}

\author{Luca Doria (on behalf of the PIENU Collaboration)}
\address{TRIUMF, 4004 Wesbrook Mall, Vancouver, BC, V6T 2A3, Canada.}
\ead{luca@triumf.ca}

\begin{abstract}
The PIENU experiment aims at the measurement of the branching ratio
$R=\Gamma (\pi\rightarrow e\nu + \pi\rightarrow e\nu\gamma)/ \Gamma (\pi\rightarrow \mu\nu + \pi\rightarrow \mu\nu\gamma)$
at the $\le 0.1\%$ precision level, with which mass scales of 1000 TeV/c$^{2}$ can be searched for new
pseudoscalar interactions. The status of the experiment is described.

\end{abstract}

\section{Motivation}
Within the Standard Model (SM), lepton universality refers to the fact that different
charged leptons have identical electroweak gauge interactions. Differences among them
can be found only in their masses and couplings to the neutral Higgs scalar, a remnant
of the mass generation through the electroweak spontaneous symmetry breaking.\\
The most stringent test of lepton universality regards the first two families 
($e-\mu$ universality) and comes from the measurement of the branching ratio
$R=\Gamma (\pi\rightarrow e\nu + \pi\rightarrow e\nu\gamma)/ \Gamma (\pi\rightarrow \mu\nu + \pi\rightarrow \mu\nu\gamma)$.
At lowest order, assuming lepton universality
\begin{equation}
R^{0}_{th} = \frac{m_{e}^{2}}{m_{\mu}^{2}}\left( \frac{m_{\pi}^{2}-m_{e}^{2}}{m_{\pi}^{2}-m_{\mu}^{2}} \right)^{2}=
1.28347 \times 10^{-4} .
\end{equation}
The smallness of $R^{0}_{th}$ is a consequence of the helicity suppression for the leptonic $\pi\rightarrow e\nu$ decay.
A measurement of this quantity includes effects of real and virtual photons and therefore
the calculation of radiative corrections is needed. The first calculation including
radiative corrections \cite{r0,r1,r2} and considering a pointlike pion gave a reduced branching ratio 
$R_{th} = 1.233 \times 10^{-4}$ with a correction of the order $\delta=(3\alpha/\pi)\ln(m_{e}/m_{\mu})$.
The latest modern calculation \cite{r3} takes into account the pion structure using chiral perturbation theory
to order $O(e^{2}p^{4})$ and radiative corrections as well:
\begin{equation}
R_{th} = (1.2352 \pm 0.0001) \times 10^{-4} .
\end{equation}
This prediction has to be compared with the present combined experimental result \cite{r5,r6}
\begin{equation}
R_{exp} = (1.230 \pm 0.004) \times 10^{-4} .
\end{equation}
The PIENU experiment aims at improving the experimental precision for $R_{exp}$ by about a factor 5,
and therefore testing the SM prediction to better than $\pm0.1\%$ level.
Such a precision experiment has the potential to uncover physics beyond the SM, measuring a deviation
from the theoretical calculations. Effects from new physics are expected in the $\Delta R / R \sim 10^{-4}-10^{-2}$
range and therefore there is a possibility to detect them or constraint the theoretical models \cite{r3}. 
There are may possible extensions of the SM which can modify R, like heavy neutrinos, extra dimensions \cite{r7},
leptoquarks \cite{r8}, compositeness \cite{r9}, charged Higgs bosons or (R-parity violating) SUSY \cite{r10}. 
Since the $\pi\rightarrow e\nu$ decay is helicity-suppressed, R is very sensitive to helicity-unsuppressed 
couplings like the pseudo-scalar ones. Pseudo-scalar contributions through interference terms are 
proportional to $1/m_{H}^{2}$,  
where $m_{H}$ is the mass of the hypothetical particle. On the contrary, lepton flavour violating 
decays such as $\mu\rightarrow e \gamma$ receive $1/m_{H}^{4}$ contributions.\\
The deviation from the SM prediction can be parameterized as \cite{r0}:
\begin{equation}
1-\frac{R_{exp}}{R_{SM}} \sim \pm \frac{\sqrt{2\pi}}{G_{\mu}} \frac{1}{\Lambda_{PS}}
\frac{m_{\pi}^{2}}{m_{e}(m_{d}+m_{u})} \approx \left( \frac{1 {\rm TeV}}{\Lambda_{PS}} \right)^{2} \times 10^{3}.
\end{equation}
ignoring the negligible contribution of the $\pi\rightarrow\mu\nu$ decay in presence of pseudoscalar interactions and assuming
a coupling similar to the weak one. The parameter $\Lambda_{PS}$ represents the mass scale of the new pseudoscalar interaction.
With the planned precision, PIENU will be sensitive to mass scales of $\Lambda_{PS} \sim O(1000 ~ {\rm TeV})$ for new pseudoscalar
interactions, which is far beyond the reach of collider experiments. Scalar couplings due to new physics can also induce changes 
to R through higher order loop corrections \cite{r10}.\\ 
The high statistics sample recorded by the PIENU experiment will also allow an improved search for massive neutrino states 
using the $\pi\rightarrow e\nu$ decay up to a mass of 130 MeV/c$^{2}$.
 
\begin{figure}[!t]
%\resizebox{\columnwidth}{!}{\includegraphics{figure0_LL.eps}}
\begin{center}
\resizebox{12cm}{!}{\includegraphics{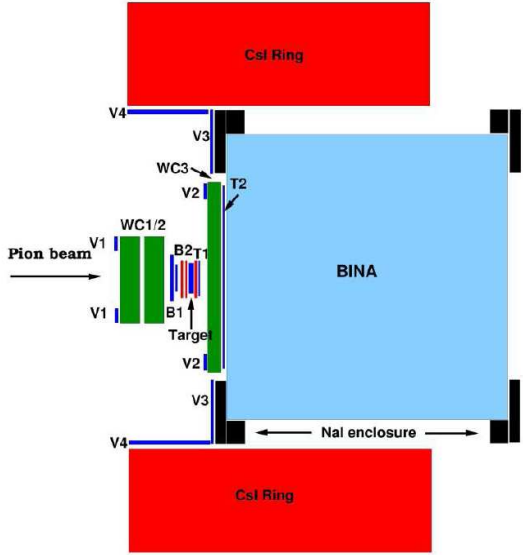}}
\end{center}
\caption{Elevation view of the PIENU setup. A large NaI(Tl) crystal (called BINA), measures the
energy of decay positrons from an active scintillator target. The NaI(Tl) crystal is surrounded
by 97 CsI crystals. The pion beam comes from the left and is tracked by wire chambers and
silicon microstrip detectors.}
\label{fig0}
\end{figure}

\begin{figure}[!t]
%\resizebox{\columnwidth}{!}{\includegraphics{figure1_LL.eps}}
\begin{center}
\resizebox{14cm}{!}{\includegraphics{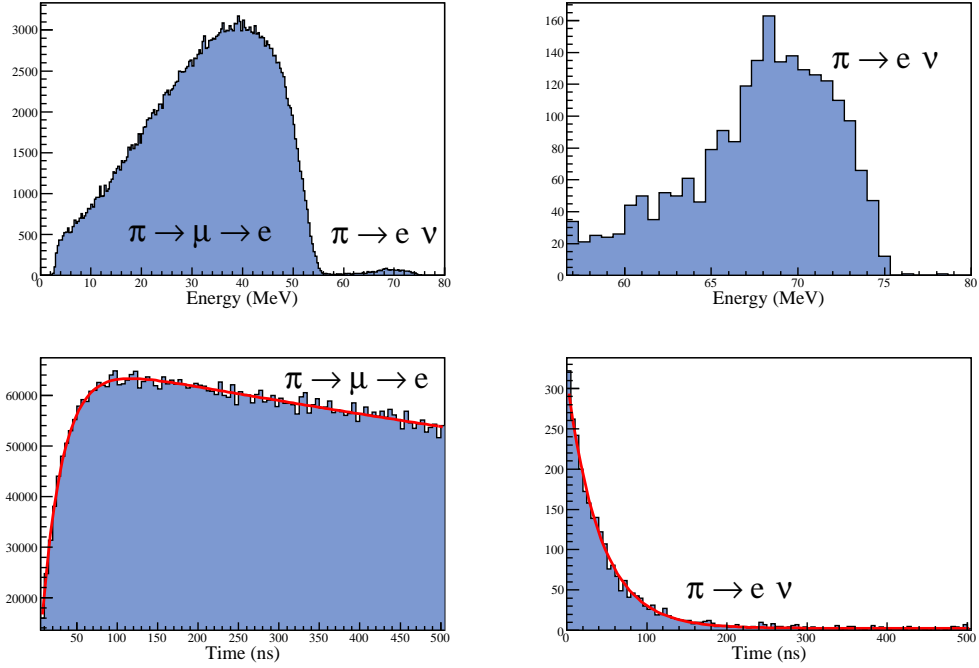}}
\end{center}
\caption{{\bf (Top Left)} Energy spectrum measured with the NaI(Tl) crystal. {\bf (Top Right)} Enlarged
energy spectrum in the $\pi\rightarrow e\nu$ region. {\bf (Bottom Left)} Time distribution of 
the $\pi^{+}\rightarrow \mu^{+} \rightarrow e^{+}$ decay chain, obtained from the low-energy region events.
{\bf (Bottom Right)} Time distribution of the $\pi^{+}\rightarrow e^{+}\nu$ decay, obtained from the high-energy
region events.}
\label{fig1}
\end{figure}

\section{The Experimental Technique}
The PIENU experiment will obtain R from the ratio of positron yields from the  $\pi^{+}\rightarrow e^{+}\nu$ decay
($E_{e^+}=69.3$ MeV) and the $\pi^{+}\rightarrow \mu^{+}\nu$ decay followed by the 
$\mu^{+} \rightarrow e^{+}\nu\bar{\nu}$ decay ($\pi^{+}\rightarrow\mu^{+}\rightarrow e^{+}$, $E_{e^+}=0-52.3$ MeV).\\
By measuring the decay positrons from both decays at the same time with the same apparatus (see fig.~\ref{fig0}), many 
normalization factors like the solid angle cancel at the first order and only small energy-dependent
effects have to be taken into account as correction.\\
The experiment is based on a stopped beam technique: a 75 MeV/c pion beam impinges on an active scintillator target
where it stops via energy loss (50-100k/s pion stop rate). 
The high-purity pion beam has a very low ($\le2\%$) positron content and it is produced 
with an energy-loss separation technique \cite{r4}.\\
Pions decay at rest into the target, which is thick enough to contain
the muons from the $\pi^{+}\rightarrow \mu^{+}\nu$ decay. 
The energy of the positrons from the target is measured with a large (48$\times$48 cm) NaI(Tl) 
crystal with 25\% solid angle and about 1\% FWHM energy resolution at 70 MeV.
The NaI(Tl) is surrounded by 97 CsI crystals for shower leakage detection.\\
Identification of pions is achieved with scintillators, wire chambers and silicon microstrip 
detectors before the target. Silicon detectors are important for identifying pion decays in flight 
before the target. After the target other scintillators and silicon detectors are used for tracking and identification
of decay positrons. Another scintillator and wire chambers are placed in front of the NaI(Tl) calorimeter
for defining the positron acceptance.
The scintillators before and after the target provide the measurement of the decay time.

\section{Analysis and Expected Precision}
The analysis strategy is based on dividing the positron energy spectrum into low-energy (containing mostly
$\pi^{+}\rightarrow \mu^{+} \rightarrow e^{+}$ decays)
and high-energy ($\pi^{+}\rightarrow e^{+}\nu$ decays) regions (see fig.~\ref{fig1}). 
The regions are divided slightly above the end-point of the $\pi^{+}\rightarrow \mu^{+} \rightarrow e^{+}$ 
spectrum (the ``Michel edge'').\\ 
The energy-based separation allows to identify the two decays, but backgrounds still remain to be 
identified and subtracted. The high energy region contains some fraction of the 
$\pi^{+}\rightarrow \mu^{+} \rightarrow e^{+}$ decays due to charged and neutral pileup effects.
The low-energy region contains a tail of $\pi^{+}\rightarrow e^{+}\nu$ decays. 
The tail originates from energy leakage
from the NaI(Tl) and radiative decays ($\pi^{+}\rightarrow e^{+}\nu\gamma$). The low-energy tail
is buried under the large amount of $\pi^{+}\rightarrow \mu^{+} \rightarrow e^{+}$ events and
its knowledge represented the major uncertainty of the previous experiments.\\
A key part of the PIENU experiment is the reduction of the low-energy tail and a
measurement of it directly from the data, without relying on
simulation. The technique to observe the low-energy tail is based on pulse-shape fitting
of the target signals with 500 MHz ADCs for identifying and suppressing the 
$\pi^{+}\rightarrow \mu^{+} \rightarrow e^{+}$ contribution. 
The improved tracking capabilities given by the combination of wire chambers
and silicon microstrip detectors will help in suppressing the remaining background from
decays in flight. A factor of 5 improvement with respect to the previous experiment \cite{r5} will
be expected for the uncertainty on the low-energy tail.
The larger solid angle will help in lowering the second largest source of uncertainty, which comes
from Coulomb and Bhabha scattering of positrons.\\ 
The extraction of the branching ratio is based on the simultaneous fit of the time spectra
of the two energy regions, taking into account the shapes of the backgrounds and the amount of the
low-energy tail.

\section{Conclusion}
The PIENU experiment is completely operational and taking physics data since 2009. With a factor 30
improvement in statistics with respect to the previous TRIUMF experiment it will be able to 
significantly reduce the main sources of uncertainty and provide the most precise measurement of the
branching ratio $R$. At PSI (Switzerland), an experiment with a similar goal is being carried out. 
Precision experiments in the LHC era can provide important and complementary tests for the Standard Model
or for New Physics scenarios.

%In the LHC era,
%the possible direct discovery of new physics signals, precision experiments are important
%for confirming and better elucidating  ??

%\begin{thebibliography}
%\bibitem{x} xx 
%\end{thebibliography}

\end{document}